\documentclass[a4paper,aps,pre,floatfix,twocolumn,showpacs,superscriptaddress,longbibliography]{revtex4-2}

\usepackage{bm}
\usepackage{graphicx}
\usepackage[caption=false]{subfig}
\usepackage[pdftex]{hyperref}

\usepackage{amssymb}
\usepackage{amsmath}
\usepackage{amsfonts}
\usepackage{color}

\begin{document}
%
%
%
%
\title{
All against the machine: the Solo score for rating skill in variable environments}
\author{David Reguera}
\affiliation{Departament de F\'isica de la Mat\`eria Condensada, Universitat de Barcelona, Mart\'i i Franqu\`es 1, E-08028 Barcelona, Spain}
\affiliation{Universitat de Barcelona Institute of Complex Systems (UBICS), Universitat de Barcelona, 08028 Barcelona, Spain}
\author{Xavier R. Hoffmann}
\affiliation{King Shared Services S.L, Ciutat de Granada 150, 08018 Barcelona, Spain}
\author{Irene P\'erez}
\author{Pol Colomer-de-Sim\'on}
\affiliation{Dribia Data Research, S.~L., Pg. de Gràcia 55, E-08007 Barcelona, Spain}

\author{Miquel Masoliver}
\author{Xavier Guardiola}
\author{Jan Wedekind}

\affiliation{King Shared Services S.L, Ciutat de Granada 150, 08018 Barcelona, Spain}

\author{Mari\'an Bogu\~n\'a}
\affiliation{Departament de F\'isica de la Mat\`eria Condensada, Universitat de Barcelona, Mart\'i i Franqu\`es 1, E-08028 Barcelona, Spain}
\affiliation{Universitat de Barcelona Institute of Complex Systems (UBICS), Universitat de Barcelona, 08028 Barcelona, Spain}

\date{\today}

\begin{abstract}
We propose a distribution-free metric to rate individual skill in ``player-versus-environment'' settings, where participants face heterogeneous tasks without direct opponents. Such settings are common in digital platforms, games, education, finance, and the benchmark evaluation of AI agents. They combine high randomness, tasks of widely varying difficulty, and unknown heterogeneity across individuals. Our metric maps each task outcome to a bounded performance score with zero population mean and variance bounded by $1/3$, regardless of the outcome distribution, so that scores are directly comparable across tasks. Aggregating these scores over tasks yields an interpretable skill score for each individual (the ``Solo'' rating), and a reshuffling null model tests whether that score exceeds what chance alone would produce. We validate the approach on the progression of $2\times10^5$ players in the mobile games Candy Crush Saga and Bubble Witch 3 Saga. The metric identifies high- and low-skill players with high statistical confidence, their classification persists over hundreds of subsequent levels, and a windowed version of the score tracks changes in performance along progression.

\end{abstract}

\maketitle

%
%
%
%
%
%

\section*{Significance statement}
Rating skill is easy when people compete against each other. It is hard when everyone competes against the same environment: students against a curriculum, investors against the market, players against a game, AI agents against a benchmark. These tasks differ in difficulty, outcomes are partly luck, and an easy success counts as much as a hard one. We show that a simple, distribution-free transform puts outcomes on all tasks on one scale, that summing it over tasks measures skill, and that a null model separates persistent skill from random luck. Tested on hundreds of thousand mobile-game players, the method offers a principled way to rank individuals, adapt challenges to ability, and compare AI systems on stochastic tasks.

\section*{Introduction}
In daily life, living organisms, including humans, face a myriad of challenges, from survival and social competition to entertainment through sports and video or table-top games. These challenges, whether for obtaining resources like food and money or for fun, vary greatly in difficulty and the required skill set, leading to diverse outcomes. This variation makes quantifying skill both complex and essential, especially in ``player-versus-environment'' (PvE) scenarios where direct competition (``player-versus-player'', PvP) is absent and randomness influences outcomes. In addition, individuals within a population differ in skill, and the challenges themselves often differ in nature and difficulty. So how can one quantify---and hence differentiate---the skill of a set of individuals based on the observed outcomes of a series of different challenges with intrinsic randomness?

This question has gained urgency with the recent rise of AI agents built on large language models. Their evaluation is a PvE problem: an agent faces a fixed set of benchmark tasks of very different, and largely unknown, difficulty, and its outcomes are random, because of sampling at non-zero temperature, non-deterministic tools and environments, and variation in how the same task is phrased. The number of attempts an agent needs to solve a task is exactly one example of the outcome variable we study here.

Historically, skill assessment has focused on direct competition, evolving from the foundational work of Zermelo~\cite{Zermelo:1929fk}, later rediscovered by Bradley and Terry~\cite{Bradley:1952qv,Davidson:1976ty}, to Elo's rating system~\cite{elo2008rating}, used to this day in chess, sports~\cite{fifa}, and games~\cite{Glickman:1999rz,LangvilleBook}. The approach has been extended to multiple types of pairwise comparisons~\cite{Newman2022}, to explicit components of luck and skill~\cite{jerdee2023luck}, and to more than two competitors at once~\cite{Herbrich:2006db}. Bradley--Terry models also underlie the arena-style evaluation of language models by pairwise human preference~\cite{Chiang:2024hl}. All these models estimate the strength of each competitor from the outcomes of contests between them. But in a PvE setting there is no opponent to rate. One can instead treat each task as an opponent with its own difficulty. This is done in practice by Elo-type ratings of chess puzzles and of exercises in adaptive learning systems~\cite{PELANEK2016169}, and more formally by item-response theory in psychometrics~\cite{LordNovick1968,Rasch1960,vanDerLindenHambleton1997}, whose Rasch model is a Bradley--Terry model between persons and items, and by knowledge tracing in education~\cite{Corbett:1994fy}. Item-response models have recently also been fitted to the benchmark results of large language models~\cite{Polo:2024zp}. All of these require a parametric link between ability, difficulty, and a binary outcome, and they estimate item parameters jointly with ability. Percentile-based scoring~\cite{HyndmanFan1996} is nonparametric but yields ranks, not a per-task score with fixed mean and controlled variance across arbitrary empirical difficulty distributions. Our goal here is different: to define a nonparametric, level-wise performance variable that is directly comparable across tasks and can be aggregated into a skill score without assuming a particular generative form for heterogeneity or noise.

In PvE scenarios, players must successfully overcome a particular challenge against a common environment. Examples range from predators hunting for prey, students completing an academic curriculum, brokers investing in the stock market, or players playing video games against the machine instead of against other players (PvP)~\cite{chick2012,yohann2010,mehta2010,beck2004,parkin2014}. Extending skill-rating systems to this case is far from trivial because the outcome of the challenge is often dominated by randomness. Additionally, different challenges may have different difficulties and dynamics, and the degree of dexterity or intellectual ability that is required to successfully overcome them varies greatly from challenge to challenge. Finally, individuals usually learn from the interaction with the environment, so their skills change over time. Accurately measuring skill in these settings requires an approach that can distinguish true skill from luck~\cite{MauboussinBook}, remain interpretable and sensitive to skill changes, and classify players into distinct levels despite the diversity of tasks. 

In this paper, we introduce the Solo score, a skill-rating metric designed to meet these requirements in PvE scenarios, named in deference to the classic single-player campaigns of video games, where one plays alone against the machine. We demonstrate its value for the case of mobile games, where a large amount of high-quality data is available. In particular, we apply it to evaluate the skill of a set of $2\times 10^5$ players of the popular game Candy Crush Saga as they progress along the different levels of the game. Our results show that the metric identifies skillful players with a high level of statistical confidence. The properties of this skill metric make it suitable to rank individuals by skill in other contexts as well, such as education or finances, and to improve the success and enjoyment of challenging or learning activities, by properly adjusting the difficulty to the skill.

Here we define the per-task performance transform, derive its properties, build the aggregated skill score and its null model, and validate them on $2\times10^5$ players, paving the way for a deeper understanding of individuals' capabilities in varied environments.

\section*{Results and Discussion}

We focus on a particular example of a mobile game: Candy Crush Saga. This popular mobile game is played by tens of millions of players every day. It poses a linear sequence of levels that players access one-by-one as the previous level is successfully completed. To pass a level, the player has to reach a specific goal, e.g. a certain score or collect a specific number of candies, by swapping two adjacent candies on a game board, using a limited number of moves and resources. The success in different levels involves different difficulties, requires different degrees of dexterity, and has a high component of randomness in the initial configuration and in the dynamics as the level gameplay unfolds. It is therefore a well-suited case study for separating luck from skill among players facing challenges of varying difficulty.  

\begin{figure}[t]
\begin{center}
\includegraphics[width=\linewidth]{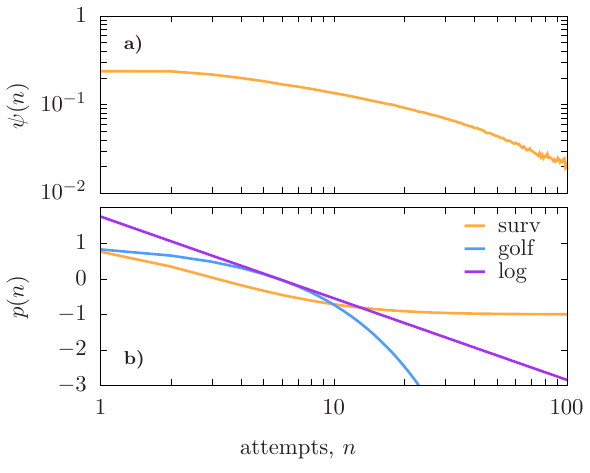}
\end{center}
\caption{(a) Distribution of pass times of level 190 of Candy Crush Saga, with average pass time $\langle n \rangle=5$. (b) Three different performance metrics of the same level: \emph{log} (violet), \emph{golf} (blue), and \emph{survival} (orange), as a function of the attempts to pass.}

\label{fig:performance_single_level}
\end{figure}

Suppose we have a population of $N_p$ players facing a set of $N_l$ challenges (hereafter levels of the game). The outcome of a player facing a particular level can be either successfully passing it or failing it. In the former case, the player starts the next new level of the game. In the latter, the player attempts to pass the level repeatedly until she succeeds. The number of attempts needed to succeed, $n$ (hereafter also called the pass time), and its distribution, $\psi(n)$, will be used to evaluate her performance in that particular level. In the case of games played sequentially following a predefined ``map'', like Candy Crush Saga, the distribution of attempts to pass a level $\psi(n)$ is, for many levels, slightly heterogeneous or even geometric. However, there are also many levels that are highly heterogeneous (see Fig.~\ref{fig:performance_single_level}(a)), with a significant number of players taking a large number of attempts to pass the level. The average number of attempts to pass a level $\langle n \rangle$ is a measure of the average difficulty of a level, and it changes for different levels. Based on this distribution, our strategy to define a skill rating metric has two steps: the first one is defining a suitable metric of individual performance in each level; in the second step, we aggregate the performance from a set of different and independent levels for defining a skill metric for each player. 

\subsection{Performance metric in a single level}
Clearly, a player that requires a small number of attempts to pass a level performs better than a player that requires a larger amount of attempts to pass the same level. Following this line of thought, there are many different possibilities to define a performance metric. The simplest option would consist in ordering players in decreasing number of pass attempts and using their rank number as performance metric. In order to obtain a bounded value, we divide the rank number by the total number of players. Then the best player would have a performance of 1, and the worst player would achieve a performance of $1/N_p \approx 0$. This metric has several issues. The first one is that it is size dependent: a player with the same number of attempts would obtain a different performance depending on the size of the group of players. Second, this ranking method needs a criterion for the resolution of ties. Third, it is not comparable between levels with different difficulties. Finally, passing on the first attempt in a very difficult level should have a higher contribution to the rating of a player's skill than passing on the first attempt in a very easy level. Yet the ranking performance would be the same in both cases. The standard evaluation of AI agents shares this flaw: success is averaged over benchmark tasks, so a trivial and a near-impossible task count equally, and repeated sampling (pass@$k$)~\cite{chen2021evaluating,yao2024taubench} addresses the randomness of outcomes but not the heterogeneity of difficulty.

\begin{figure}[t]
\begin{center}
\includegraphics[width=\linewidth]{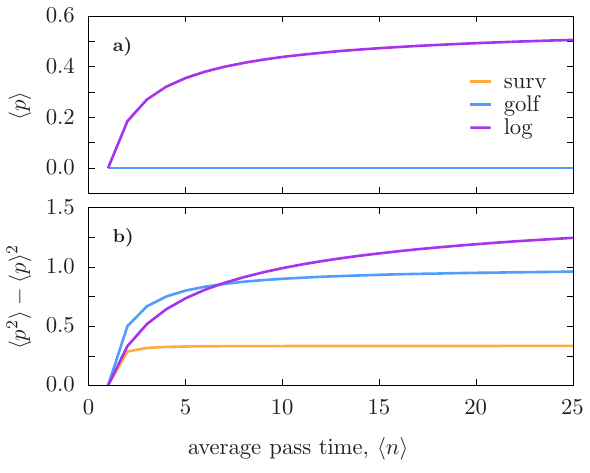}
\end{center}
\caption{Average (a) and variance (b) of the \emph{golf} (blue), \emph{log} (violet), and \emph{survival} (orange) performance metrics in random levels exponentially distributed, as a function of the average pass time $\langle n \rangle$.}
\label{fig:performance_many_levels}
\end{figure}

One option to fix these problems is using z-scores, {\it i.e.}, computing the difference with respect to the average number of attempts of the set of players and normalize by the standard deviation. Consider a level with average number of attempts to pass $\langle n \rangle$ and standard deviation $\sigma_n=\sqrt{\langle n^2\rangle-\langle n \rangle^2}$. Then a player that passed with $n$ attempts would have a performance of $p=(\langle n\rangle -n)/\sigma_n$. This metric is suitable for comparing levels of different difficulty. But the assumption of normality is very far from the fat-tailed pass distributions that we observed in the empirical data. 

In a similar approach---inspired by the scoring system used in golf---we can normalize by the average number of attempts instead of its standard deviation. With this \emph{golf} metric, the previous player would have performance $p^{\text{golf}}(n)=(\langle n\rangle -n)/\langle n\rangle$. This metric has an upper bound of $1$ (see Fig.~\ref{fig:performance_single_level}(b)) and the elegant property that the average over all players in a level is always 0 (see Fig.~\ref{fig:performance_many_levels}(a)). However, given the power-law-like shape of the attempt distribution, the \emph{golf} performance of players that require a very large amount of attempts in order to pass the level have very large magnitude (but negative sign). As a consequence, a skillful player that would perform badly in a level would be heavily penalized in its skill rating.

An alternative to mitigate the problem of outliers is the use of logarithms. We define the \emph{log} performance as $p^{\text{log}}(n)=-\log(n/\langle n\rangle)$, so that positive and negative values correspond, respectively, to good and bad performances. Although extreme values for large numbers of attempts are reduced, its maximum value, $\log\langle n\rangle$ at $n=1$, grows without bound with the difficulty of the level (see Fig.~\ref{fig:performance_single_level}(b)). In addition, the average over players does not have a controlled value and depends on the particularities of the pass distribution. Specifically, its population mean is $\log\langle n\rangle-\langle\log n\rangle$, which is non-negative by Jensen's inequality and depends on the shape of $\psi(n)$. Moreover, if $\psi(n)$ decays as a power law with exponent at most 2, $\langle n\rangle$ diverges and neither $p^{\text{log}}$ nor $p^{\text{golf}}$ is defined.

We therefore propose a performance metric that fixes all these issues. Consider a level for which the number of attempts required to pass is a discrete random variable with the probability mass function $\psi(n)$. Define the (upper-tail) survival function $\Psi(n)=\sum_{n'=n}^{\infty}\psi(n')$, {\it i.e.}, the probability that passing requires at least $n$ attempts. We define the \emph{survival} performance of a player with $n$ attempts as
\begin{equation}
p^{\text{sur}}(n)=2\Psi(n)-\psi(n)-1.
\label{eq:performance}
\end{equation}
This definition is a monotone rescaling of the survival probability: it assigns high performance to outcomes that are rare in the upper tail of $\psi(n)$ (few attempts), and it standardizes scores across levels with different $\psi(n)$ by construction~\footnote{Up to sign and scale, Eq.~\eqref{eq:performance} is the mid-distribution function of Parzen~\cite{Parzen2004,MaGentonParzen2011}, $F_{\text{mid}}(n)=\mathrm{Pr}(N\le n)-\tfrac{1}{2}\psi(n)$, since $p^{\text{sur}}(n)=1-2F_{\text{mid}}(n)$; to our knowledge, its use as a cross-task standardizer for skill rating is new. }. As a scoring metric, Eq.~\eqref{eq:performance} has several desirable properties for a skill-rating metric:
\begin{itemize}
\item[i)] It is a monotonically decreasing function of $n$ and is bounded between $-1$ and $1$.
\item[ii)] If all players clear a level in the same number of attempts, they cannot be distinguished. In this case, Eq.~\eqref{eq:performance} assigns every player a performance of $0$.
\item[iii)] For very difficult levels, players who clear the level in the first attempts attain values close to $1$ (specifically, $p^{\text{sur}}(1)=1-\psi(1)$), thus providing a clear separation between skilled and unskilled players.
\item[iv)] Its mean over the player population is $0$, and its variance is bounded above by $1/3$ for any distribution $\psi(n)$; see Fig.~\ref{fig:performance_many_levels}(b). These moment constraints make the scale of $p^{\mathrm{sur}}$ comparable across levels, which is crucial when aggregating performance over tasks with widely different difficulty distributions.
\end{itemize}
Properties ii) and iv) follow in closed form from the mid-distribution representation: for any $\psi(n)$, $\langle p^{\text{sur}}\rangle=0$ and
\begin{equation}
\sigma^2_{p^{\text{sur}}}=\frac{1}{3}\Big(1-\sum_{n}\psi(n)^3\Big),
\label{eq:var_psur}
\end{equation}
so the variance vanishes for a point-mass distribution, is bounded by $1/3$, and reaches the bound only in the continuous limit where all $\psi(n)\to 0$.

A similar approach can also be used to rank the performance of individuals based on the outcomes of continuous processes. Suppose that performance is quantified by a variable $x \in (-\infty,\infty)$ with probability density $p(x)$ and complementary cumulative distribution function $P_c(x)=\int_x^\infty p(x')\,dx'$. If better performance corresponds to larger values of $x$, then we define the survival performance as
\begin{equation}
p^{\text{sur}}(x)=1-2P_c(x).
\end{equation}
In this case, $p^{\text{sur}}(x)$ is uniformly distributed on $(-1,1)$, with zero mean and variance $1/3$. Indeed, by the probability integral transform, $P_c(x)$ is uniformly distributed on $(0,1)$ for any continuous $x$~\cite{CasellaBerger2002}.

Now that we have a reliable metric to quantify a player's performance on a single level, we can quantify the consistency of a player's performance relative to the rest of the population. Note that aggregation does not require independence of outcomes across levels; it only requires that the level-wise mapping be comparable across levels, which Eq.~\eqref{eq:performance} guarantees. Selection effects remain: the set of players observed at deeper levels is conditioned on earlier success. We therefore complement aggregation with a reshuffling-based null model that preserves each level's empirical difficulty while destroying cross-level correlations attributable to persistent skill.

\subsection{Aggregated skill score}

The properties of $p^{\text{sur}}(n)$ explained above are fundamental for defining a good skill rating metric. If the distribution of attempts of players to pass a level $\psi(n)$ is the result of some random process that is independent of players' skill, then---by virtue of the central limit theorem---the aggregation of this performance metric over several levels yields a normal distribution with zero mean and a small variance, which can be used as a good reference to distinguish true skill from pure luck. Instead, if $\psi(n)$ is the combination of the properties of the level and the skill of players, then skillful players will perform systematically better whereas players with poor skills will perform worse level after level.

\begin{figure}[t]
\begin{center}
\includegraphics[width=\linewidth]{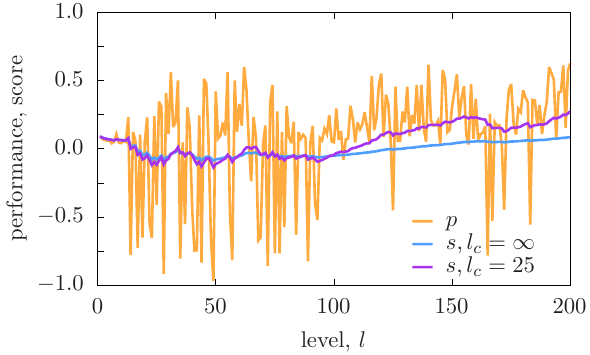}
\end{center}
\caption{The orange line shows the instantaneous \emph{survival} performance of an arbitrary player during the first 200 levels of Candy Crush Saga. The violet and blue curves show the Solo skill score computed from Eq.~\eqref{eq:skillmetric} with $l_c=25$ and $l_c=\infty$, respectively.}
\label{fig:skill_average}
\end{figure}

In Fig.~\ref{fig:skill_average}, we show $p^{\text{sur}}(n)$ of a randomly selected player during the first 200 levels of Candy Crush Saga. As can be seen, the signal is very noisy, with sudden high and low peaks, so that skill cannot be quantified by a single level. Instead, we must aggregate the performance over a certain amount of levels to get a stable measure. To do that, let us first sort levels in chronological order $l=1,\cdots,N_l$. Suppose that player $i$ has a sequence of attempts to pass the levels given by $\{n_{i,l}\; ; \; l=1,\cdots,N_l \}$. Then, we compute the skill score of player $i$ on level $l\le N_l$, the Solo score, as
\begin{equation}
s_i(l)=\sum_{l'=1}^l w_{l,l'}p^{\text{sur}}(n_{i,l'}),
\label{eq:skillmetric}
\end{equation}
where $p^{\text{sur}}(n_{i,l})$ is the performance of player $i$ on level $l$ given by Eq.~\eqref{eq:performance} and the weight $w_{l,l'}$ measures the importance that the players' performance on level $l'$ has on the player's score at level $l$. The specific choice of these weights is arbitrary, although different choices produce similar results. Here, we define the weights as 
\begin{equation}
w_{l,l'}=\frac{1-e^{-1/l_c}}{1-e^{-l/l_c}}e^{-(l-l')/l_c}\; , \; l'\le l,
\end{equation}
where $l_c$ is a characteristic window size. Notice that with this choice, weights are normalized, that is, $\sum_{l'=1}^l w_{l,l'}=1$. With this particular form, recently played levels have a higher contribution to the skill metric whereas levels that were played more than $l_c$ levels in the past have a vanishing effect. Setting $l_c =\infty$ produces a cumulative skill metric that weights all played levels equally up to $l$, so it will be a very stable metric to characterize the global skill of players. Setting $l_c$ to a finite value defines a skill metric that is sensitive to variations of players' skill due to learning or deterioration.

Fig.~\ref{fig:skill_average} compares the two choices, showing the instantaneous performance of a player of Candy Crush Saga during the first 200 levels and the skill metric Eq.~\eqref{eq:skillmetric} with $l_c =\infty $ and $l_c=25$. The former is more robust and fluctuates less than the latter. The latter, however, is more suitable to capture rapid variations along the progression. In particular, the metric with $l_c=25$ correctly differentiates two different regimes: the player starts with low skill up to level 100 and, from that moment on she exhibits a persistent positive performance. On the other hand, the cumulative score is less sensitive to these variations because it is an average of the performance over the entire player's history.

\subsection{Persistent player performance}

The observations from our pool of players could be the result of them playing a fully random game, which would render skill an irrelevant property. If this null assumption were correct, the aggregated skill metric Eq.~\eqref{eq:skillmetric} would be a Gaussian with zero mean and its variance bounded by
\begin{equation}
\sigma_{s_i(l)}^2 \le \frac{1}{3} \frac{(1-e^{-1/l_c})(1+e^{-l/l_c})}{(1+e^{-1/l_c})(1-e^{-l/l_c})} \approx \frac{1}{3l} \text{  for  } l_c \gg 1.
\label{eq:sigma}
\end{equation}
We can use this Gaussian distribution to define a null model able to distinguish sheer luck from true skill. Specifically, fix a level depth $l$ and select the $N_p(l)$ players who have played all levels $1,\dots,l$. For each level $l'\le l$, randomly permute the observed attempt counts $\{n_{i,l'}\}_{i=1}^{N_p(l)}$ across players, thereby preserving the empirical distribution $\psi_{l'}(n)$ at that level while destroying player-specific correlations across levels. We then recompute $p^{\text{sur}}_{i,l'}$ and the skill score for each player. Repeating this procedure yields the null distribution against which empirical skill scores and quantiles can be compared. 

Alternatively, a closed-form estimate follows from Eq.~\eqref{eq:var_psur}. Under the null model, the variance of $s_i(l)$ is exactly $\sum_{l'\le l} w_{l,l'}^2\,\sigma^2_{l'}$, with $\sigma^2_{l'}$ given by Eq.~\eqref{eq:var_psur} evaluated on the empirical $\psi_{l'}(n)$. If attempts are geometrically distributed, $\psi(n)=(1-q)q^{n-1}$, then $\sum_n\psi(n)^3=1/(3\langle n\rangle^2-3\langle n\rangle+1)$ and this reduces to
\begin{equation}
\sigma_{s_i(l)}^2 \approx \sum_{l'=1}^l w_{l,l'}^2 \frac{\langle n_{l'} \rangle \left[\langle n_{l'} \rangle-1\right]}{1-3 \langle n_{l'} \rangle+3\langle n_{l'} \rangle^2},
\label{eq:sigma_approx}
\end{equation}
where $\langle n_{l} \rangle$ is the average number of attempts at level $l$. This estimate avoids computationally costly simulations and can be used to set the limit that differentiates true skill from pure luck.

\begin{figure}[t]
\begin{center}
\includegraphics[width=\linewidth]{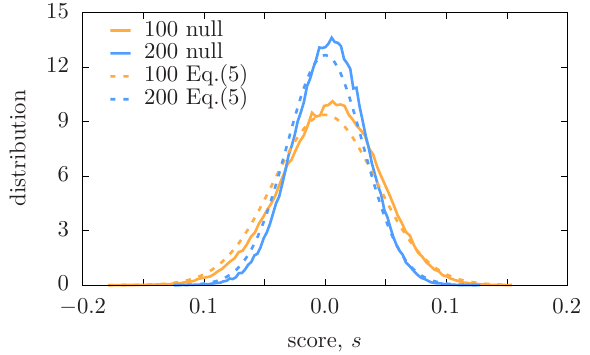}
\end{center}
\caption{Skill score distribution of simulated null players (solid) and the Gaussian approximation given by Eq.~\eqref{eq:sigma_approx} (dashed) on levels 100 and 200 for Candy Crush Saga. Skill score is measured with Eq.~\eqref{eq:skillmetric} and $l_c=\infty$.}
\label{fig:null_model}
\end{figure}

Figure~\ref{fig:null_model} shows the distribution of skill scores from the null model at levels 100 and 200 of Candy Crush Saga for a cohort of $N_p=2\times 10^5$ players using Eq.~\eqref{eq:skillmetric} with $l_c=\infty$. For this null model, we observe that the skill metric is close to being Gaussian distributed, and the spread decreases for larger levels as expected. Figure~\ref{fig:null_model} also shows the Gaussian distributions with variances given by Eq.~\eqref{eq:sigma_approx}.

\begin{figure}[t]
\begin{center}
\includegraphics[width=\linewidth]{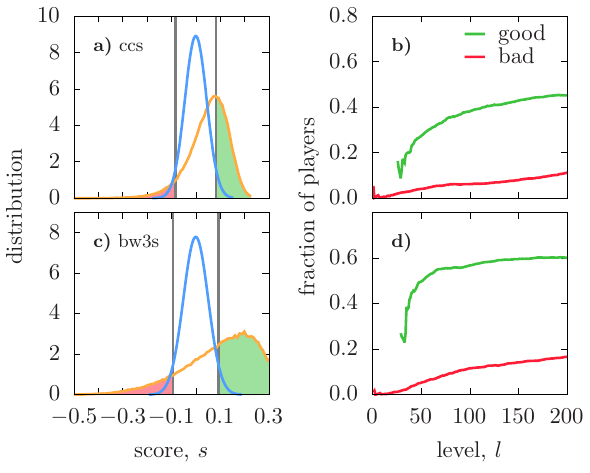}
\end{center}
\caption{Skill score distribution of null (blue) and empirical (orange) players on level 100 of Candy Crush Saga (a) and Bubble Witch 3 Saga (c). (b,d) Fraction of empirical players that are classified as good (green) and bad (red) on levels 1 to 200 of Candy Crush Saga and Bubble Witch 3 Saga.}
\label{fig:quantiles}

\end{figure}

Figure~\ref{fig:quantiles}(a) shows the null and empirical skill distributions at level 100. As can be clearly seen, the empirical distribution is significantly different from the one derived from the null model, indicating that players' skill has a major contribution, even for a game like Candy Crush Saga, where success has a very strong random component. The difference between the null and the empirical distribution gets accentuated for other types of games with higher difficulty or requiring more dexterity, like Bubble Witch 3 Saga (Figure~\ref{fig:quantiles}(c)), which is a bubble shooter puzzle game. 

The null model allows us to classify players as ``good'' (high skill) or ``bad'' (low skill). At a 95\% confidence level, we compute from the null distribution the lower and upper quantiles that correspond, respectively, to the worst and best 2.5\% of null players. These two thresholds are then applied to the empirical players. By construction, only 2.5\% of null players fall beyond each threshold, so the test is informative to the extent that the empirical fractions exceed this value.

Comparing each player's actual score to these two thresholds we can classify them as good (their score is higher than the upper quantile), bad (their score is lower than the lower quantile), or neutral (their score lies in between the two quantiles). We see that a substantial fraction of players are classified as good in terms of skill. In fact, from figure~\ref{fig:quantiles}(b,d) we observe that the fraction of good players grows with the number of levels, as does the fraction of bad players (albeit less rapidly). This behavior is partially due to the shrinking spread of the null distribution, allowing a better differentiation of skill. Accumulating performance over many levels allows a finer segmentation, ranking players by skill rather than only classifying them as good or bad.

\begin{figure}[t]
\begin{center}
\includegraphics[width=\linewidth]{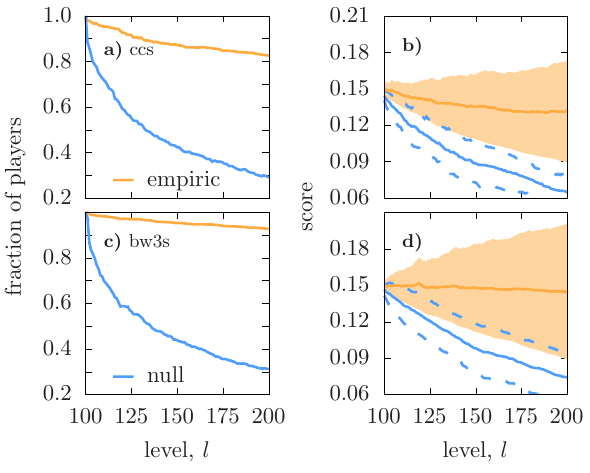}
\end{center}
\caption{(a,c) Fraction of the empirical (orange) and null (blue) players that are classified as good on level 100 of Candy Crush Saga and Bubble Witch 3 Saga and remain classified as good in the following levels. (b,d) Average score and standard deviation of empirical (orange line and shaded area) and null (blue solid and dashed lines) players that are classified as good on level 100 with score $s(100)  \approx 0.15$ of Candy Crush Saga and of Bubble Witch 3 Saga.}
\label{fig:diffusion}
\end{figure}

We still need to check for the consistency of our segmentation of players as good, bad, or neutral. We start by selecting all the players that are classified as good on level 100. We then track them in further levels and compute the fraction of this subset of players that remain classified as good in subsequent levels. Figure~\ref{fig:diffusion}(a,c) shows the evolution of this fraction until level 200, both for the empirical dataset and the null model. We do observe a reduction, which means that some players' skill is getting worse, but this decline is much less drastic than in the null model. To complete the analysis, we select all the players with similar skill score on level 100, in particular between 0.14 and 0.16. We then track them again in further levels and compute their average score, as shown in figure~\ref{fig:diffusion}(b,d). The score of these players remains almost constant. This contrasts starkly with the null model, where, naturally, players show a rapid regression to the mean. For the null model with $l_c=\infty$ this regression is exactly what we would expect, $\langle s(l)\rangle = 100\,s(100)/l$, which gives $0.075$ at $l=200$. The difference with respect to this regression to the mean is even more accentuated for games with more difficult challenges or requiring higher dexterity, like Bubble Witch 3 Saga. Taken together, these results indicate that our segmentation of players is  robust.

\section*{Conclusion} \label{sec:concluding_remarks}

We have introduced the Solo score, a metric for rating individual skill in variable environments, using Candy Crush Saga and Bubble Witch 3 Saga as examples. The \emph{survival} performance transforms each outcome into a score with zero mean and bounded variance for any outcome distribution, so that performance on tasks of very different difficulty can be compared on one scale. Aggregated over tasks, it yields a skill score that meets the requirements set out in the Introduction: a null model separates persistent skill from luck; the score is interpretable; a windowed version is sensitive to changes in skill along progression; and it classifies players into groups whose membership persists over hundreds of subsequent levels, even in a game where success has a strong random component.

The same construction applies whenever performance is measured as a count (attempts, failures before success) or as a continuous outcome (via the complementary CDF transform). Examples include estimating learning progress across heterogeneous exercises, assessing trader performance across assets with different volatility profiles, and, beyond ranking, adapting the difficulty of a challenge to an individual's ability to improve engagement and learning. A limitation is that comparisons are conditioned on the observed cohort (e.g., players who reach a given depth), so population-level statements require care. Still, the method provides a principled and reproducible way to compare individuals under strong heterogeneity without specifying a parametric noise model.

An application we consider especially timely is the evaluation of AI agents. Benchmark evaluation is a PvE setting: tasks vary widely in difficulty, outcomes are random, and the number of attempts until success is the natural outcome variable. The \emph{survival} performance applies to it directly, with the population of agents (or of agent runs) playing the role of the player population, and the reshuffling null model tests whether an agent's advantage on a leaderboard exceeds what chance alone would produce, a question current leaderboards rarely address~\cite{Miller:2024qf}. Two points deserve study: agent populations are small compared with the player population numbers we used in this study, so $\psi(n)$ must be estimated from fewer samples; and scores are relative to the population evaluated, as in any rating system. We regard the application of this framework to public benchmark results as a natural and worthwhile next step.

\bibliographystyle{naturemag}
\bibliography{skill}


\section*{Competing interests}

The authors declare no competing interests.

%

%
  
%
%
%
%


%
%

%
\end{document}